\documentclass[%
 reprint,
superscriptaddress,
 amsmath,amssymb,
 aps,nofootinbib
]{revtex4-1}
\usepackage{dsfont}
\usepackage{graphicx}
\usepackage{dcolumn}
\usepackage{bm}
\usepackage{color}
\usepackage{epsfig}
\usepackage[colorlinks=true
,urlcolor=blue
,anchorcolor=blue
,citecolor=blue
,filecolor=blue
,linkcolor=blue
,menucolor=blue
,linktocpage=true
,pdfproducer=medialab
,pdfa=true
]{hyperref}
\usepackage{natbib}
\usepackage{float}
\usepackage{footmisc}
\usepackage{slashed}
\usepackage{comment}

\usepackage{orcidlink}

\definecolor{lightred}{rgb}{1,0.5,0.5}
\definecolor{lightgreen}{rgb}{0.5,1,0.5}
\definecolor{lightblue}{rgb}{0.5,0.5,1}
\definecolor{lightcyan}{rgb}{0.5,0.75,0.75}
\definecolor{lightmagenta}{rgb}{0.75,0.5,0.75}
\definecolor{customgreen}{rgb}{0.494,1,0.502}

\newcommand{\eV}{\mathinner{\mathrm{eV}}}

\newcommand{\MeV}{\mathinner{\mathrm{MeV}}}
\newcommand{\GeV}{\mathinner{\mathrm{GeV}}}

\begin{document}

\title{Thermal damping of spectator scalar dark matter:\\ 
Relaxing the overclosure and  isocurvature bounds}

\author{Erwin H.~Tanin \orcidlink{0000-0002-3204-0969}}
\email{ehtanin@berkeley.edu}
\affiliation{Department of Physics, University of California Berkeley}
\affiliation{School of Physics and Astronomy, University of Minnesota}
\affiliation{Leinweber Institute for Theoretical Physics at Stanford, Department of Physics, Stanford University, Stanford, California 94305, USA}

\author{Sarunas Verner \orcidlink{0000-0003-4870-0826}}
\email{verner@uchicago.edu}
\affiliation{Kavli Institute for Cosmological Physics, University of Chicago, 5640 South Ellis Ave.,
Chicago, IL 60637, USA}

\begin{abstract}
A light scalar present during inflation generically acquires a misalignment and can later constitute the dark matter. Since it fluctuates independently of the inflaton, however, it carries isocurvature perturbations. Isocurvature and overclosure bounds together rule out the simplest version of this scenario at high inflationary scales. We show that a weak coupling to the primordial plasma can relax both bounds simultaneously. The plasma depletes the scalar abundance through thermal damping, a dissipative effect arising from the plasma's finite equilibration time, thereby relaxing the overclosure bound. If the damping grows with the oscillation amplitude, it can also drive different initial amplitudes onto a common attractor that depends only on the local radiation temperature. This synchronizes the scalar with the adiabatic clock, thus suppressing its isocurvature. The resulting isocurvature bound on the inflationary scale relaxes by exactly the factor by which damping reduces the abundance, in addition to the relaxation from the larger initial misalignment now allowed. This result holds for any damping rate quadratic in the amplitude, regardless of its temperature dependence.
\end{abstract}

\maketitle

\section{Introduction}

Scalar fields are common in extensions of the Standard Model (SM) \cite{Essig:2013lka,Antypas:2022asj,Graham:2025gtd}. A scalar lighter than the inflationary Hubble scale $H_{\rm inf}$ undergoes de Sitter fluctuations with amplitude $\delta\chi\sim H_{\rm inf}/2\pi$ and therefore generically acquires a misalignment $X_i$ by the end of inflation \cite{Starobinsky:1994bd,Kolb:2023ydq,Graham:2018jyp,Hertzberg:2008wr,Kainulainen:2016vzv}. At late times, the resulting scalar condensate may constitute all or part of the dark matter (DM), with an abundance set by $X_i$ \cite{Kolb:2023ydq,Markkanen:2018gcw,Tenkanen:2019aij}. Since a spectator scalar fluctuates independently of the inflaton, its density perturbations contain an isocurvature component of relative amplitude $\delta\rho_\chi/\rho_\chi\sim (H_{\rm inf}/2\pi)/ X_i$, which is subject to stringent cosmic microwave background (CMB) limits \cite{Planck:2018jri}.

This leads to a tension. Avoiding overclosure favors a smaller initial misalignment $X_i$, whereas suppressing the isocurvature favors a larger one \cite{Chung:2004nh,Linde:1991km,Graham:2018jyp}. Together, they bound the scalar mass and $H_{\rm inf}$, strongly restricting the viable parameter space \cite{Tenkanen:2019aij}. These conclusions, however, rest on the assumption that the spectator scalar evolves in isolation from the primordial plasma after inflation.

If the scalar couples, even weakly, to particles populated by reheating, its post-inflation evolution can change dramatically. The plasma modifies the scalar dynamics in two qualitatively distinct ways. First, equilibrium thermal corrections to the effective potential, such as a thermal mass, change the onset of scalar oscillations and how the condensate subsequently redshifts \cite{Henrich:2026kcl,Bellac:2011kqa,Quiros:1999jp}. Second, the plasma responds to the evolving scalar with a finite relaxation time. The resulting lag behind instantaneous equilibrium generates a dissipative response that drains energy from the scalar field and heats the plasma \cite{Tanin:2017bzm,Banerjee:2025qyg,Banerjee:2025nvs,Hosoya:1983id,Yokoyama:1998ju,Yokoyama:2004pf,Mukaida:2012qn,Mukaida:2012bz,Mukaida:2013xxa,Mukaida:2014yia,Bastero-Gil:2010dgy,Bodeker:2022ihg,Drewes:2013iaa}. This thermal damping is not captured by the equilibrium effective potential and can substantially alter both the abundance and the perturbations of the scalar.

In this work we study how these thermal effects reshape the background and perturbation evolutions of a spectator scalar. We illustrate the mechanism in a minimal model in which a real scalar $\chi$ controls the mass of a fermion $\psi$ that is in thermal contact with the SM plasma. The thermal population of $\psi$ generates both a thermal effective potential and thermal damping for $\chi$. Although the detailed outcomes are model-dependent, the mechanism applies broadly whenever an oscillating scalar modulates the properties of a species in the thermal bath.

At the background level, thermal damping depletes the scalar condensate and thereby relaxes the usual overclosure bound. Moreover, under suitable conditions, thermal damping can drive the scalar amplitude toward a dynamical attractor on which the thermal-damping rate is comparable to the Hubble rate \cite{Banerjee:2025nvs,Banerjee:2025qyg}. A range of initial misalignments then converges onto a common evolution, which the field tracks until it eventually decouples. The relic abundance is consequently set by where the attractor tracking ends rather than by the initial spectator misalignment $X_i$. In some sense, the abundance-setting mechanism resembles thermal freeze-out, but for a coherent condensate that is never thermalized.

At the perturbation level, the same attractor suppresses pre-existing isocurvature. While the scalar is on the attractor, its amplitude is dictated by the local plasma temperature and no longer provides an independent clock. A perturbation along the attractor is therefore adiabatic, \textit{i.e.}, equivalent to a local time shift  \cite{Weinberg:2003sw,Allali:2025pja}. Isocurvature perturbations, which correspond to displacements away from the attractor, are thermally damped during the attractor tracking. A small residual isocurvature may survive if the tracking is too brief to erase the  primordial spectator mode completely.

Thermal damping can thus relax the overclosure and isocurvature constraints simultaneously, removing the usual tension between them. Furthermore, once the attractor is reached, the relic abundance becomes largely insensitive to the unknown inflationary initial conditions and is instead controlled by the low-energy interactions and late thermal history, which are more accessible to experiment.

This paper is organized as follows. In Section~\ref{s:background} we introduce the model, review the thermal effects, discuss the thermal-damping attractor, and define the scope of the effective description. In Section~\ref{s:perturbation} we study the evolution of superhorizon perturbations and the suppression of primordial isocurvature. In Section~\ref{s:tension} we show how thermal damping relaxes both the overclosure and isocurvature bounds. We conclude in Section~\ref{s:conclusion}. The appendices contain the derivation of the perturbation equations (Appendix~\ref{a:perturbationderivation}), the relation between the isocurvature mode used in the main text and the scalar-radiation relative entropy (Appendix~\ref{a:relativeisocurvature}), a discussion of subleading isocurvature sources (Appendix~\ref{a:subleadingisocurvature}), and model realizations of various thermalization rates (Appendix~\ref{a:models}).

\section{Spectator scalar dark matter coupled to a thermal bath}
\label{s:background}

We consider a real spectator scalar $\chi$ coupled to a Dirac fermion $\psi$ that exchanges energy with the SM plasma. A minimal effective description is
\begin{align}
    \mathcal{L}\supset -V_{\rm vac}(\chi)-(m_\psi-g\chi)\bar{\psi}\psi+\mathcal{L}_{\psi\leftrightarrow\text{SM}}\,,
\end{align}
where $V_{\rm vac}$ is the vacuum potential of $\chi$ and $\mathcal{L}_{\psi\leftrightarrow\text{SM}}$ contains interactions that thermally equilibrate $\psi$ with the SM. The scalar $\chi$ modulates the effective fermion $\psi$ mass,
\begin{align}
    M_\psi(\chi)=|m_\psi-g\chi|\,.
\end{align}
We focus on a post-reheating epoch during which the fermion $\psi$ is relativistic and remains efficiently thermalized with the SM bath at temperature $T$, with an effective thermalization rate
    \begin{align}
        \Gamma_{\rm th}=\kappa T^n\, ,
    \end{align}
where the coefficient $\kappa$ has mass dimension $1-n$ and the exponent $n$ depends on the operators in $\mathcal{L}_{\psi\leftrightarrow\text{SM}}$; we discuss representative portals in Appendix~\ref{a:models}. The rate $\Gamma_{\rm th}$ describes how quickly the $\psi$ distribution relaxes toward its instantaneous equilibrium as $M_\psi(\chi)$ evolves. We assume that $\mathcal{L}_{\psi\leftrightarrow\text{SM}}$ does not have further relevant effects beyond setting $\Gamma_{\rm th}$.

\subsection{Thermal Potential and Damping}

The equation of motion of the scalar field $\chi$ is
\begin{align}
    \Box\chi+V_{\rm vac}^\prime(\chi)=g\left<\bar{\psi}\psi\right>\,. \label{eq:eom}
\end{align}
Here and throughout this paper we use the mostly minus metric signature, $(+---)$. If the fermion $\psi$ adjusted instantaneously to the scalar, its response $\left<\bar{\psi}\psi\right>$ would depend only on the instantaneous values of $\chi$ and $T$. Its effect could then be absorbed into the effective potential of $\chi$. In reality, the $\psi$ distribution relaxes toward equilibrium at the finite rate $\Gamma_{\rm th}$. Its response lags slightly behind the evolving $\chi$ and has a component proportional to $\dot{\chi}$, which gives rise to thermal damping.

When $\psi$ remains close to thermal equilibrium, the right-hand side of \eqref{eq:eom} can be evaluated as \cite{Banerjee:2025nvs,Banerjee:2025qyg}
\begin{align}
    g\left<\bar{\psi}\psi\right>=m_{\chi,\rm th}^2\left(\frac{m_\psi}{g}-\chi\right)-\Upsilon\dot{\chi},\,\,\,\,\, m_{\chi,\rm th}=c_mgT.
\end{align}
The first term is the equilibrium response, characterized by the thermal mass $m_{\chi,\rm th}$, where $c_m=1/\sqrt{6}$ for a Dirac fermion $\psi$. The second term is the non-equilibrium lag giving rise to thermal damping with coefficient $\Upsilon$. For the homogeneous background $\bar{\chi}$, the equation of motion reduces to
\begin{align}
     \ddot{\bar{\chi}}+(3H+\bar{\Upsilon})\dot{\bar{\chi}}+M_\chi^2\left(\bar{\chi}-\chi_{\rm th}\right)&=0 \,,
\end{align}
with $M_\chi^2=V_{\rm vac}''+m_{\chi,\rm th}^2$ the effective mass of $\chi$, $\chi_{\rm th}=(m_{\chi,\rm th}^2/M_\chi^2)m_\psi/g$ the $\psi$-induced shift of the effective potential minimum, and $\bar{\Upsilon}$ the thermal damping coefficient evaluated on the background $\bar{\chi}$.

We assume $\chi$ has a large amplitude $\gg m_\psi/g$ (which also implies $\gg\chi_{\rm th}$) and oscillates rapidly compared to thermal and Hubble damping ($M_\chi\gg\Upsilon, H$).\footnote{The case with amplitude $\lesssim m_\psi/g$ is studied in \cite{Batell:2021ofv,Batell:2022qvr,Batell:2026sml,Graham:2025gtd}.} The background $\bar{\chi}$ is then approximately sinusoidal
\begin{align}
    \bar\chi=X(t)\cos\theta(t),\quad \dot{\theta}(t)\approx M_\chi(t)\,, \label{eq:backgroundapprox}
\end{align}
with a slowly varying envelope $X$, $|\dot{X}|/X\ll M_{\chi}$, that evolves as \cite{Banerjee:2025nvs,Banerjee:2025qyg}
\begin{align}
    \frac{\dot{X}}{X}=-\frac{1}{2}\left(3H+\tilde{\Upsilon}+\frac{\dot{M}_\chi}{M_\chi}\right),\quad \tilde{\Upsilon}\equiv \frac{\left<\bar{\Upsilon}\dot{\bar{\chi}}^2\right>}{\left<\dot{\bar{\chi}}^2\right>}\,. \label{eq:envelope}
\end{align}
Here $\left<\ldots\right>$ denotes an average over an oscillation period, and $\tilde{\Upsilon}$ is the $\dot{\bar{\chi}}^2$-weighted oscillation-averaged damping rate. In this regime, the $\tilde{\Upsilon}$ takes the form \cite{Banerjee:2025nvs,Banerjee:2025qyg}
\begin{align}
    \tilde{\Upsilon}=c_\Upsilon\frac{g^4 X^2}{\Gamma_{\rm th}}\frac{\Gamma_{\rm th}^2}{\Gamma_{\rm th}^2+4M_\chi^2}\,,\label{eq:Upsilontilde}
\end{align}
with $c_\Upsilon=3c_m^2/2\pi^2$ a dimensionless coefficient determined by the number of degrees of freedom of $\psi$.

In what follows we focus, for simplicity, on an epoch during radiation domination, with approximately constant relativistic degrees of freedom, in which the potential of $\chi$ is dominated by its thermal mass:
\begin{align}
    \frac{\dot{T}}{T}\approx -H,\quad M_\chi\approx m_{\chi,\rm th}=c_mg T\,.
\end{align}
Then we can use the temperature $T$ as a clock and rewrite the envelope equation, Eq.~\eqref{eq:envelope}, as
\begin{align}
    \xi=\xi_{\rm free}+\frac{\tilde{\Upsilon}}{2H}\,, \label{eq:envelope2}
\end{align}
where we defined the logarithmic slope of the envelope
\begin{align}
     \xi\equiv\frac{d\ln X}{d\ln T},\quad \left.\xi_{\rm free}\equiv\xi\right|_{\tilde{\Upsilon}=0}=1\,.\label{eq:xi}
\end{align}
Without damping, the amplitude scales as $X\propto T^{\xi_{\rm free}}=T$ under thermal mass. Thermal damping increases the exponent by $\tilde{\Upsilon}/2H$.

\subsection{Thermal-Damping Attractor}

Since thermal damping weakens as the amplitude decreases, $\tilde{\Upsilon}\propto X^2$, it is self-limiting. When $X$ is sufficiently large that $\tilde{\Upsilon}\gg H$, $X$ decreases rapidly, and $\tilde{\Upsilon}$ drops to $\sim H$ in a fraction of a Hubble time \cite{Banerjee:2025nvs,Banerjee:2025qyg}. Whether it then remains at $\tilde{\Upsilon}\sim H$ or continues to drop toward $\tilde{\Upsilon}\ll H$ depends on the temperature dependence of $\tilde{\Upsilon}$ at fixed amplitude, which is controlled by $\Gamma_{\rm th}(T)$ and $M_\chi(T)$.

To see which occurs, consider the amplitude $X_\Upsilon(T)$ at which $\tilde{\Upsilon}=c_{\rm track}H$ for some constant $c_{\rm track}$. From Eq.~\eqref{eq:Upsilontilde},
\begin{align}
    X_\Upsilon(T)=\frac{\sqrt{c_{\rm track}H\Gamma_{\rm th}}}{g^2\sqrt{c_\Upsilon}}\sqrt{\frac{\Gamma_{\rm th}^2+4M_\chi^2}{\Gamma_{\rm th}^2}}\,. \label{eq:XUps}
\end{align}
Differentiating with respect to $T$ at fixed $c_{\rm track}$ and using $H\propto T^2$ gives its slope
\begin{align}
    \xi_\Upsilon\equiv\frac{d\ln X_\Upsilon}{d\ln T}=1-\frac{1}{2}\left.\frac{\partial\ln\tilde{\Upsilon}}{\partial\ln T}\right|_X\,. \label{eq:xiups}   
\end{align}
Requiring $X=X_\Upsilon$ to solve the envelope equation, Eq.~\eqref{eq:envelope2}, fixes $c_{\rm track}=2(\xi_\Upsilon-\xi_{\rm free})$, and so
\begin{align}
    c_{\rm track}=-\left.\frac{\partial \ln\tilde{\Upsilon}}{\partial\ln T}\right|_{X}=n-\frac{8(n-1)M_\chi^2}{\Gamma_{\rm th}^2+4M_\chi^2}\,.\label{eq:ctrack}
\end{align}
The amplitude $X_\Upsilon\propto T^{1+c_{\rm track}/2}$ acts as a dynamical attractor if the undamped amplitude $X\propto T$ decreases more slowly than $X_\Upsilon$ as $T$ goes down, \textit{i.e.}, if
\begin{align}
    &c_{\rm track}>0 &&(X_\Upsilon\text{ is an attractor})\,.
\end{align}
An amplitude above $X_\Upsilon$ is damped faster than the Hubble rate and is driven down to it, while an amplitude below $X_\Upsilon$ catches up with it, so $X_\Upsilon$ is an attractor from both sides.\footnote{Note that the amplitude dependence of $\Upsilon$ is essential for the existence of the attractor $X_\Upsilon$. The sphaleron friction arising from an axion-like coupling $aG\tilde{G}$, for example, is independent of the amplitude and does not produce an attractor of this kind \cite{McLerran:1990de,Berghaus:2019whh,Ji:2021mvg,Papageorgiou:2022prc, Choi:2022nlt}.} In the fast thermalization ($\Gamma_{\rm th}\gg M_\chi$) and slow thermalization ($\Gamma_{\rm th}\ll M_\chi$) regimes, $c_{\rm track}$ simplifies to
\begin{align}
    c_{\rm track}&=\begin{cases}
        n, &\Gamma_{\rm th}\gg M_\chi\\
        2-n, &\Gamma_{\rm th}\ll M_\chi
    \end{cases}\,.
\end{align}
So $X_\Upsilon$ is an attractor for $n>0$ in the fast-thermalization regime and for $n<2$ in the slow-thermalization regime.

Our description applies after reheating, once $\psi$ is thermalized and $\chi$ has started to oscillate. The approach to the attractor depends on where the amplitude $X$ starts in this regime, which is set by inflation and reheating. If $X$ starts above $X_\Upsilon$, it drops to $X_{\Upsilon}$ within a fraction of a Hubble time. If $X$ starts below $X_\Upsilon$, it redshifts freely as $X\propto T$, while $X_\Upsilon\propto T^{1+c_{\rm track}/2}$. The ratio $X/X_{\Upsilon}$ therefore grows by $e^{c_{\rm track}/2}$ per $e$-fold, and the two meet after $N_{\rm approach}\approx (2/c_{\rm track})\left.\ln(X_\Upsilon/X)\right|_{T_{\rm start}}$ $e$-folds, where $T_{\rm start}$ is the temperature at which our description first applies. We denote the temperatures at the start and end of tracking by $T_{\rm in}$ and $T_{\rm out}$, and the number of $e$-folds of tracking by
\begin{align}
    N_{\rm track}\equiv \ln\frac{T_{\rm in}}{T_{\rm out}}\,. \label{eq:Ntrack}
\end{align}

As stated earlier, we work in the regime where the thermal mass dominates, the amplitude is large $X\gg m_\psi/g$, and the exponent $n$ is fixed. Continued tracking requires $c_{\rm track}>0$ and that $\psi$ remains relativistic and thermalized, $T\gtrsim \text{max}(m_\psi,gX)$ and $\Gamma_{\rm th}\gtrsim H$. Tracking ends when one of these conditions ceases to hold. Along the attractor, the relativistic condition $T\gtrsim \text{max}(m_\psi,gX)$ is less strict compared to $X\gg m_\psi/g$. The thermalization requirement $\Gamma_{\rm th}\gtrsim H$ can be violated with time only for the $n>2$ case where $\Gamma_{\rm th}/H\propto T^{n-2}$ decreases. In that case, however, $c_{\rm track}$ changes sign first, because at that point $\Gamma_{\rm th}\sim M_\chi\gg H$. The tracking therefore ends on its own only through $c_{\rm track}$ changing sign. Setting Eq.~\eqref{eq:ctrack} to zero gives
\begin{align}
    T_{\rm out}&= \left(\frac{2c_mg}{\kappa}\sqrt{1-\frac{2}{n}}\right)^{\frac{1}{n-1}},\quad\text{for } n> 2\,.\label{eq:Tout}
\end{align}
This endpoint is relevant only for $n>2$, because only then does $c_{\rm track}$ change sign as $\Gamma_{\rm th}/M_\chi$ evolves. 

For $0<n\leq 2$, $c_{\rm track}$ never changes sign on its own. When $n=2$, $\Gamma_{\rm th}/M_\chi$ decreases as the temperature decreases, and $c_{\rm track}=2\Gamma_{\rm th}^2/(\Gamma_{\rm th}^2+4M_\chi^2)$ decreases with it but stays positive. For $n=1$, $\Gamma_{\rm th}/M_\chi$ is constant and $c_{\rm track}=1$. Tracking thus does not end on its own for $0<n\leq 2$, and it can persist for many $e$-folds. It can instead end at a temperature $T_*$ where $n$ changes in a way such that $c_{\rm track}<0$. We discuss an example of this in Section~\ref{s:tension}. 

Our analysis must be extended if one of its assumptions fails before tracking ends, namely if the amplitude drops to $X\sim m_\psi/g$, if the thermal mass falls below the bare mass, or if $\psi$ becomes non-relativistic. We restrict our analysis to the parameter space where tracking ends first.

\subsection{The Scope of This Work}

A complete theory would specify the origin of the scalar background and its perturbations, their evolution through reheating, and the later dynamics that determine the relic abundance and observable perturbations, possibly with different degrees of freedom and effective descriptions at different stages of this evolution. We instead restrict our analysis to a finite window from a few $e$-folds before the onset of attractor tracking at $T_{\rm in}$ to the end of tracking at $T_{\rm out}$. Throughout this interval, we assume that the thermal effects described above dominate the dynamics of the scalar $\chi$. We do not require this minimal description, or even the same degrees of freedom, to hold during inflation or after $T_{\rm out}$. This lets us study the thermal mechanism without committing to a single effective theory valid from inflation to the present day. 

Within this window, we assume that the EFT remains valid up to a cutoff $\Lambda$ above $T_{\rm in}$ and that the scalar mass is protected from $\sim g\Lambda$ mass corrections. The Coleman-Weinberg potential generated by $\psi$, $\Delta V_{\rm CW}(\chi)=-[M_\psi^4(\chi)/16\pi^2][\ln M_\psi^2(\chi)/\mu^2-3/2]$ is negligible compared to the thermal potential as long as $gX\ll T$, i.e., as long as $\psi$ is relativistic. After $T_{\rm out}$, this potential and other vacuum contributions, such as the bare mass, may come to dominate. 

We take the scalar background and its superhorizon perturbations shortly before $T_{\rm in}$ as initial data, without specifying their earlier origin. As long as the initial conditions lie within the basin of attraction and tracking lasts sufficiently long, different initial conditions converge toward a common temperature-dependent evolution, and memory of the earlier evolution is erased up to a small residual. Our robust background result is that thermal damping decreases the scalar amplitude relative to the undamped evolution. As we show in the next section, convergence onto the thermal-damping attractor also suppresses isocurvature. Once suppressed, isocurvature cannot be regenerated on superhorizon scales by later evolution.

\section{Superhorizon Perturbations}
\label{s:perturbation}

\subsection{Evolution of Superhorizon Perturbations}
BBN requires $\psi$, and hence the thermal effects on $\chi$, to be absent at the latest by $T\sim \MeV$. Hence, modes with $k\lesssim \left.aH\right|_{T\sim \MeV}\sim 10^4\text{ Mpc}^{-1}$ are superhorizon during the entire phase in which thermal effects are significant.\footnote{ This includes all CMB-observable and large-scale-structure modes.} The equation of motion for superhorizon $\chi$ perturbations $\delta\chi$ is
\begin{align}
    \delta\ddot{\chi}+(3H+\bar{\Upsilon})\delta\dot{\chi}+\left(M_\chi^2+\frac{d\bar{\Upsilon}}{d\bar{\chi}}\dot{\bar{\chi}}\right)\delta \chi&=S\,. \label{eq:perturbationevolution}
\end{align}
For $X\gg m_\psi/g$, the source is $S=S_\Phi+S_{\rm eq}+S_\Upsilon$ with
\begin{align}
    S_\Phi&=4\dot{\Phi}\dot{\bar{\chi}}-2\Phi \left(M_\chi^2\bar{\chi}+\frac{1}{2}{\bar{\Upsilon}\dot{\bar{\chi}}}\right)\,,\\
    S_{\rm eq}&=-2\frac{\delta T}{T}m_{\chi,\rm th}^2\bar{\chi}\,,\\
    S_\Upsilon&=-\frac{\delta T}{T}\left.\frac{\partial\bar{\Upsilon}}{\partial\ln T}\right|_{\bar{\chi}}\dot{\bar{\chi}}\,.
\end{align}
$S_\Phi$,  $S_{\rm eq}$, and $S_\Upsilon$ arise, respectively, from gravitational driving by the metric perturbation ($\propto \Phi,\dot{\Phi}$), the temperature dependence of the (equilibrium) thermal effective potential, and the temperature dependence of thermal damping. We provide a derivation of the perturbation equation in Appendix~\ref{a:perturbationderivation}. By the fluctuation-dissipation relation, thermal damping should in general be accompanied with a stochastic thermal noise term in the equation of motion \cite{Bastero-Gil:2014jsa}. As we discuss in Appendix~\ref{a:subleadingisocurvature}, this noise has a blue spectrum and is negligible for the \textit{superhorizon} modes considered here, so we omit it from Eq.~\eqref{eq:perturbationevolution}.

\subsection{Adiabatic and Isocurvature Modes}

The adiabatic mode of $\delta\chi$ corresponds to a local time shift $t\rightarrow t+\delta t$ of the background solution $\bar{\chi}$ (Eq.~\eqref{eq:backgroundapprox})
\begin{align}
    \delta\chi_{\rm ad}&=\dot{\bar{\chi}}\delta t=X\left(-M_\chi\sin\theta+\frac{\dot{X}}{X}\cos\theta\right)\delta t\,. \label{eq:dchiad}
\end{align}
The first and second terms stem from phase and amplitude perturbations, respectively. In the limit $(|\dot X|/X)/M_\chi\ll1$, the adiabatic mode $\delta\chi_{\rm ad}$ is predominantly in the phase direction.

Any perturbation of $\chi$ that is not of the form of Eq.~\eqref{eq:dchiad} is isocurvature. For $(|\dot X|/X)/M_\chi\ll1$, the isocurvature mode is naively just the amplitude perturbation $\delta X$. However, Eq.~\eqref{eq:dchiad} shows that a part $\dot{X}\delta t=X \xi\delta T/T$ of the amplitude perturbation, with $\xi$ defined in Eq.~\eqref{eq:xi}, is itself adiabatic. Subtracting this adiabatic contamination from the amplitude perturbation, we find that the fractional amplitude perturbation not accounted for by the local radiation clock is
\begin{align}
    I\equiv\frac{\delta X}{X}-\xi \frac{\delta T}{T}\,, \label{eq:Idefinition}
\end{align}
which is gauge-invariant and vanishes identically for a pure time shift. The isocurvature variable $I$ can be understood as the fractional amplitude perturbation on uniform temperature slices, $I=\left.\delta X/X\right|_{\delta T=0}$. To connect to observables, one can convert $I$ into the relative isocurvature between $\chi$ and radiation (see Appendix~\ref{a:relativeisocurvature}),
\begin{align}
    S_{\chi r}=-3H\left(\frac{\delta\rho_\chi}{\dot{\rho_\chi}}-\frac{\delta\rho_{\rm rad}}{\dot{\rho}_{\rm rad}}\right)=\frac{3 I}{1+\xi}\,,
\end{align}
when the thermal mass dominates the scalar potential. This confirms that $I=0$ is equivalent to vanishing relative entropy between $\chi$ and radiation.

\subsection{Erasure of Isocurvature}

The time evolution of $I$ follows from time averaging Eq.~\eqref{eq:perturbationevolution}. For $\bar{\Upsilon}\propto X^2$ and $M_\chi\approx m_{\chi,\rm th}$ one obtains (see Appendix~\ref{a:perturbationderivation} for a derivation)
\begin{align}
    \dot{I}+\frac{1}{2}\left.\frac{\partial\tilde{\Upsilon}}{\partial\ln X}\right|_{T} I=-[\dot{\xi}+\tilde{\Upsilon}(\xi-\xi_\Upsilon)]\frac{\delta T}{T}\,, \label{eq:Ievolution}
\end{align}
with $\xi_{\Upsilon}$ defined in Eq.~\eqref{eq:xiups}. The above equation implies that $I$ is damped at the rate $(1/2)\left.\partial\tilde{\Upsilon}/\partial\ln X\right|_T=\tilde{\Upsilon}$, for $\tilde{\Upsilon}$ as given in Eq.~\eqref{eq:Upsilontilde}. Note that the amplitude dependence of the thermal damping, $\tilde{\Upsilon}\propto X^2$, is crucial for the damping of $I$. It can be checked with Eq.~\eqref{eq:envelope2} and \eqref{eq:xiups} that the source term vanishes identically on any thermal-mass-dominated background solution,
\begin{align}
    \dot{\xi}+\tilde{\Upsilon}(\xi-\xi_\Upsilon)=0\,.
\end{align}
Thus, an adiabatic perturbation with $I=0$ remains adiabatic, as expected from Weinberg's adiabatic theorem \cite{Weinberg:2003sw}. The variable $I$ is consequently damped as 
\begin{align}
    I(t)=I(t_i) e^{-\int_{t_i}^{t} dt'\tilde{\Upsilon}(t')} \,.\label{eq:isocurvatureerasure}
\end{align}
This holds during the approach to the attractor, during tracking, and after decoupling, when $\tilde{\Upsilon}/H\rightarrow 0$ and $I$ freezes. Thus the same thermal damping that depletes the background $\chi$ also erases $I$. In fact,  $I$ is suppressed by exactly the same factor as the squared background amplitude $X^2$ and hence the abundance $\Omega_\chi\propto X^2$ at a given time (temperature)
\begin{align}
    \frac{I}{I(t_i)}=\frac{X^2}{X_{\rm free}^2}=\frac{\Omega_\chi}{\Omega_\chi^{\rm free}}\,,\label{eq:IratioXratio}
\end{align}
where $X_{\rm free}=X_{\Upsilon=0}$ is the amplitude that $\chi$ would have in the absence of thermal damping, starting from the same $X(t_i)$.

Isocurvature is present when $\chi$ perturbations carry information beyond a single clock. This could arise, for instance, from the inflationary spectator initial condition $\delta X_i\sim H_{\rm inf}/2\pi$. After reheating, the universe then contains two local clocks: the thermal bath temperature $T$ and the scalar amplitude $X$. Since the thermal damping rate, and hence the energy exchange between $\chi$ and the thermal bath, is a function of both $X$ and $T$, $\tilde{\Upsilon}(X,T)$, it can drive patches with the same temperature $T$ but different scalar amplitudes $X$ toward one another. Since $\tilde{\Upsilon}\propto X^2$, patches with larger amplitudes are damped faster, and their amplitudes converge toward a common value. The role of the attractor is to sustain $\tilde{\Upsilon}\sim H$ for many $e$-folds, so that the suppression in Eq.~\eqref{eq:isocurvatureerasure} accumulates. After many $e$-folds of attractor tracking, the amplitude becomes so close to $X_\Upsilon(T)$ that it is essentially a function of $T$ alone. The system is then effectively single-clock.

A nonzero residual isocurvature can nevertheless survive if tracking is too short to erase the primordial perturbation. In addition, there are at least two effects not captured by the separate-universe picture that may source isocurvature at subleading order. First, gradient terms generate a correlated isocurvature suppressed by $(k/aH)^2$. Second, thermal fluctuations, expected from the fluctuation-dissipation relation, generate uncorrelated white noise, \textit{i.e.} a dimensionless power spectrum $\propto k^3$, on scales larger than the thermal correlation length $\sim T^{-1}$. We discuss these further in Appendix~\ref{a:subleadingisocurvature}. Both subleading effects are negligible on CMB scales, but could serve as interesting targets for future probes of small-scale structures. The perturbation variable $I$ does not describe all possible departures from a common time shift. Perturbations of the oscillation phase $\theta$ of the form $\delta\theta_{\rm iso}\equiv\delta\theta-M_\chi\delta t$ are not thermally damped (see Appendix~\ref{a:perturbationderivation}). However, small phase perturbations do not contribute at linear order to the oscillation-averaged energy density $\left<\rho_\chi\right>=M_\chi^2 X^2/2$, and hence do not source energy-density isocurvature.

\begin{figure*}[ht!]
    \centering
    \includegraphics[width=0.9\linewidth]{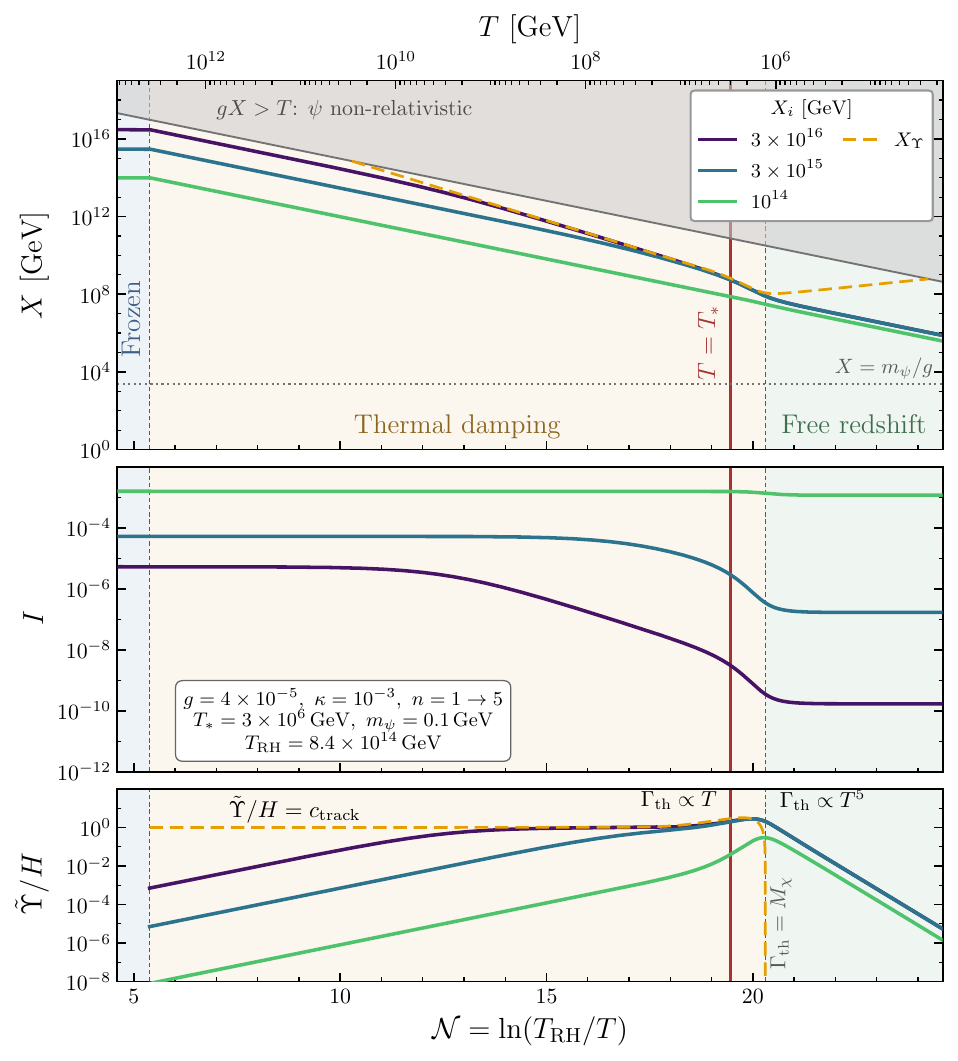}
    \caption{Evolution of the oscillation amplitude $X$ (top), the fractional amplitude perturbation on isothermal slices as a proxy for isocurvature $I\equiv \delta X/X-(d\ln X/d\ln T)\delta T/T$ (middle), and the oscillation-averaged thermal damping rate to Hubble rate ratio $\tilde\Upsilon/H$ (bottom). Here, $g=4\times10^{-5}$, $\kappa=10^{-3}$, $T_*=3\times10^{6}\,{\rm GeV}$, $m_\psi=0.1\,{\rm GeV}$, and $H_{\rm inf}=10^{12}\,{\rm GeV}$, corresponding to $T_{\rm RH}=8.4\times10^{14}\,{\rm GeV}$ for instant reheating. We use Eq.~(\ref{eq:thermalization}) and solve Eqs.~(\ref{eq:envelope}) and (\ref{eq:isocurvatureerasure}) with $I_i=H_{\rm inf}/(2\pi X_i)$. The field remains frozen until $M_\chi=3H$ at $T_{\rm osc}=4\times 10^{12}\GeV$. In the top panel, the orange dashed curve is $X_\Upsilon$ from Eq.~(\ref{eq:XUps}) with $c_{\rm track}$ replaced by $n={\rm d}\ln\Gamma_{\rm th}/{\rm d}\ln T$ and in the bottom panel, it is $c_{\rm track}$ from Eq.~(\ref{eq:ctrack}). In the gray region, $gX>T$, $\psi$ is non-relativistic. The two larger $X_i$ redshift freely, $X\propto T$, until they reach $X_\Upsilon$ and then track it, while the smallest $X_i$ is only slightly thermally damped. Tracking ends at $T_{\rm out}=1.3\times10^{6}\,{\rm GeV}$, where $c_{\rm track}=0$. Afterwards $\tilde\Upsilon/H$ falls as $T^3$ and $I$ freezes, suppressed by ($3\times10^{-5}$, $3\times10^{-3}$, $0.74$)  for the three $X_i$ ($3\times 10^{16}\GeV,3\times 10^{15}\GeV, 10^{14}\GeV$), the same factors by which their abundances $\Omega_\chi\propto X^2$ are depleted relative to the free evolution. Throughout, $X\gg m_\psi/g$ (dotted).}
    \label{fig:XvsT}
\end{figure*}

\section{Thermal damping relaxes the overclosure-isocurvature tension}
\label{s:tension}

\subsection{The standard tension}
Consider first spectator scalar dark matter without thermal effects. If the scalar is light during inflation, $m_\chi\ll H_{\rm inf}$, it acquires isocurvature perturbation uncorrelated with the adiabatic mode, with power spectrum $P_{\rm iso}\approx 4\left(H_{\rm inf}/2\pi{X_i}\right)^2$. The Planck constraint on uncorrelated isocurvature is commonly expressed as $P_{\rm iso}/(P_{\mathcal{R}}+P_{\rm iso})<0.04$ at the usual CMB pivot scale, with $\sqrt{P_{\mathcal{R}}}\approx 5\times 10^{-5}$~\cite{Planck:2018jri}. This simplifies to
\begin{align}
    X_i\gtrsim 10^{12}\GeV\left(\frac{H_{\rm inf}}{10^{7}\GeV}\right)\,.
\end{align}
In the absence of thermal effects, avoiding overclosure imposes $X_i\lesssim (2\rho_{\rm DM,0}/m_\chi^2)^{1/2}(g_*T_{\rm osc}^3/g_{*,0}T_0^3)^{1/2}$, where $T_{\rm osc}$ is the temperature at which $H=m_\chi$, which amounts to
\begin{align}
    X_i\lesssim 10^{12}\GeV\left(\frac{m_\chi}{\eV}\right)^{-1/4}\,.
\end{align}
Together, the two constraints leave no viable parameter space for  $H_{\rm inf}\gtrsim 10^7\GeV\left(m_\chi/\eV\right)^{-1/4}$. In the rest of this section we follow the isocurvature evolution for a concrete benchmark and show that thermal damping can revive this parameter space.

In Fig.~\ref{fig:XvsT} we plot $X$, $I$, and $\tilde{\Upsilon}/H$ for $g =4\times  10^{-5}$, $\kappa=10^{-3}$, $n=1$ ($n=5$) at $T\gtrsim T_*$ ($T\lesssim T_*$) with $T_*=3\times 10^6\GeV$, $m_\psi=0.1\,\mathrm{GeV}$.  We take an inflationary Hubble scale $H_{\rm inf}=10^{12} \,\mathrm{GeV}$, and assume instantaneous reheating with a reheating temperature $T_{\rm RH}=8.4\times10^{14} \, \mathrm{GeV}$. We display the evolution for three initial amplitudes $X_i ={3\times 10^{16},3\times10^{15},10^{14}}\,\mathrm{GeV}$. The trajectories with the two larger $X_i$ merge into a single curve, and consequently their relic abundance is independent of $X_i$. Isocurvature is suppressed during attractor tracking and remains constant after tracking ends at $T_{\rm out}=1.3\times 10^6\GeV$.

\subsection{A benchmark}
We take tracking to occur while the thermalization rate $\Gamma_{\rm th}=\kappa T^n$ scales with $n=1$. In this case, $\Gamma_{\rm th}/m_{\chi,\rm th}$ is temperature independent and the tracking does not end by itself. We end it by letting the exponent switch at a temperature $T_*$ to $n=5$, for which tracking ends by itself when $\Gamma_{\rm th}$ falls below $\sim M_\chi$: 
\begin{align}
    \label{eq:thermalization}
    \Gamma_{\rm th}=\begin{cases}
    \kappa T, &T\gtrsim T_*\\
    \displaystyle \kappa T_*\left(\frac{T}{T_*}\right)^{5}, &T\lesssim T_* 
    \end{cases}\,.
\end{align}
This arises naturally, for instance, if $\psi$ interacts with the SM plasma via a vector mediator $V$ with mass $m_V\sim T_*$ and SM coupling $\alpha_V$. For $T\gg m_V$ the mediator is effectively massless and $\Gamma_{\rm th}\sim \alpha_V^2T$ and for $T\ll m_V$ the mediator can be integrated out and $\Gamma_{\rm th}\sim \alpha_V^2T^5/m_V^4$. This structure is the same as for SM neutrinos, whose weak interaction rate changes from $\sim \alpha_W^2T$ to $\sim G_F^2T^5$ at around the electroweak scale. 

Right below $T_*$, the ratio $\Gamma_{\rm th}/M_\chi$ is  $\kappa/(c_m g)$. If $\kappa/(c_mg)\ll 1$, then tracking ends immediately at $T_*$. Otherwise, $\Gamma_{\rm th}/M_\chi$ decreases as $T^4$ until it falls below unity. Accounting for both cases, tracking ends at
\begin{align}
    T_{\rm out}\sim T_*\text{ min}\left[1,\left(\sqrt{\frac{12}{5}}\frac{c_m g}{\kappa}\right)^{1/4}\right]\,.
\end{align}
For the benchmark of Fig.~\ref{fig:XvsT}, $\kappa/(c_mg)\approx 60$, giving $T_{\rm out}\approx 0.4T_*=1.2\times 10^6\GeV$, consistent with the figure.

\subsection{Isocurvature evolution}

The evolution of the isocurvature can be divided into four stages: generation of spectator fluctuations during inflation, onset of oscillations and approach to the attractor, attractor tracking, and late-time free evolution after departure from the attractor. Throughout the last three stages, Eq.~\eqref{eq:isocurvatureerasure} applies, so the net suppression is given by $e^{-\int dt\,\tilde{\Upsilon}}$. The stages differ only in how the integral accumulates.

During inflation, $\chi$ field acquires fluctuations $\delta X\sim H_{\rm inf}/2\pi$, giving an initial fractional field perturbation $\delta X_i/X_i\sim H_{\rm inf}/(2\pi X_i)$ uncorrelated with the adiabatic (inflaton) perturbation. Since the radiation bath is not yet present, the isocurvature mode $I$ defined in Eq.~\eqref{eq:Idefinition} in terms of the radiation temperature is not applicable. Nevertheless, it is clear that the inflaton and the spectator $\chi$ provide two independent clocks, and $\delta X_i$ is an isocurvature perturbation. The initial condition for $I$ for the subsequent evolution is therefore $I_i=\delta X_i/X_i$.

After reheating, the thermal population of $\psi$ generates a thermal mass and thermal damping of $\chi$. The $\chi$ field begins to oscillate at the temperature $T_{\rm osc}$ where $M_\chi\sim H$. If at that time $X_i$ is well above $X_\Upsilon$ then $X$ is driven rapidly onto the attractor $X_\Upsilon$ within a fraction of a Hubble time at $T_{\rm osc}$. During the approach, the isocurvature amplitude decays as $I\propto X^2$, as per Eq.~\eqref{eq:IratioXratio}. By the end of the approach, $X\approx X_\Upsilon(T)$, and the initial spectator value $I_i$ has decayed to $I(T_{\rm in})\approx I_i(X_\Upsilon/X_i)_{T_{\rm osc}}^2$. If instead $X$ starts below $X_\Upsilon$, damping is negligible until the two meet after $N_{\rm approach}\approx (2/c_{\rm track})\left.\ln(X_\Upsilon/X)\right|_{T_{\rm osc}}$ $e$-folds (see Section~\ref{s:background}), and $I(T_{\rm in})=I_i$. Every trajectory plotted in Fig.~\ref{fig:XvsT} starts below the attractor.

Once on the attractor, $\tilde{\Upsilon}=c_{\rm track}H$, and $I$ falls exponentially in the number of $e$-folds as $e^{-\int_{T_{\rm out}}^{T_{\rm in}} d\ln T\,c_{\rm track}}$, which is $T_{\rm out}/T_{\rm in}$ for $c_{\rm track}=1$. Deep into the tracking, the evolution of $\chi$ is dictated by $T$ alone. The two initially independent clocks, $T$ and $X$, have been driven toward each other, and the system is effectively single-clock. Unless the tracking period is short, the relative isocurvature between scalar and radiation $S_{\chi r}\propto I$ is essentially erased. At the end of tracking, $X$ stops chasing $X_\Upsilon$, $\tilde{\Upsilon}/H$ drops below unity, and thermal damping shuts off. Thereafter, $I$, and with it $S_{\chi r}$, is constant. The prediction of our scenario is therefore the standard spectator isocurvature mode uncorrelated with the adiabatic perturbation, but with its amplitude suppressed by a calculable factor,
\begin{align}
   \epsilon\equiv\frac{I(T_{\rm out})}{I_i}\approx \frac{X_\Upsilon^2(T_{\rm out})}{[X_i (T_{\rm out}/T_{\rm osc})]^2} \,,
\end{align}
which follows from Eq.~\eqref{eq:IratioXratio}, since $X_{\rm free}(T_{\rm out)}=X_iT_{\rm out}/T_{\rm osc}$. For $c_m g/\kappa\ll 1$, as is the case for the benchmark of Fig.~\ref{fig:XvsT}, this boils down to $\epsilon\approx 0.2 \kappa^{1/4}g^{-7/4}M_{\rm pl}T_*/X_i^2$. Thus the isocurvature is more suppressed for a larger $X_i$, as reflected in Fig.~\ref{fig:XvsT}.

Since the suppression is the same for every mode that is superhorizon at $T_{\rm out}$ ($k$-independent), the original spectral shape is retained and only the normalization changes. Even without an attractor, any epoch with $\tilde{\Upsilon}\gtrsim H$ suppresses the isocurvature by $e^{-\int dt\tilde{\Upsilon}}$. The attractor makes this suppression large and the final abundance largely insensitive to $X_i$.

\subsection{Relaxing the tension}
By Eq.~\eqref{eq:IratioXratio}, the $\chi$ energy density $\rho_\chi=M_\chi^2X^2/2$ is suppressed by the same factor $\epsilon$, relative to undamped evolution from the same $X_i$, as the isocurvature. Thermal mass alone already raises $T_{\rm osc}$ and can reduce $\rho_{\chi,0}\propto X_i^2/(M_{\rm pl}T_{\rm osc})$ relative to the pure bare-mass misalignment case. Thermal damping depletes $\chi$ further and relaxes the overclosure constraint even more. Due to the attractor, the relic abundance is now independent of $X_i$ and $H_{\rm inf}$, and is instead set by where tracking ends. Since $P_{\rm iso}\propto \epsilon^2$, the isocurvature limit relaxes to $H_{\rm inf}\lesssim \epsilon^{-1}\times3\times10^7\GeV(X_i/10^{12}\GeV)$ if $\chi$ constitutes the full DM. The tension is thus relaxed and the parameter space at high $H_{\rm inf}$ is reopened.

As found in Eq.~\eqref{eq:IratioXratio}, the isocurvature suppression factor $\epsilon\equiv I_{\rm out}/I_i$ is also the abundance depletion factor $\epsilon =\Omega_{\rm DM}/\Omega_{\chi}^{\rm free}(X_i)$, where $\Omega_{\chi}^{\rm free}(X_i)=\Omega_{\chi}^{\Upsilon=0}(X_i)$ is the abundance the same initial amplitude would produce without thermal damping ($\Upsilon=0$), and we again assume that $\chi$ constitutes the DM at late times. One can thus rewrite the relaxed isocurvature bound in a form that is largely independent of the details of the thermal damping\footnote{This result is specific to $\tilde{\Upsilon}\propto X^2$. If it instead scales as $\tilde{\Upsilon}\propto X^p$, the bound would relax as $\propto [\Omega_{\chi}^{\rm free}(X_i)/\Omega_{\rm DM}]^{p/2}$.}
\begin{align}
    H_{\rm inf}\lesssim 3\times 10^7\GeV\left(\frac{X_i}{10^{12}\GeV}\right)\left(\frac{\Omega_{\chi}^{\rm free}(X_i)}{\Omega_{\rm DM}}\right)\,. \label{eq:punchline}
\end{align}
Thus, the isocurvature bound is relaxed by the factor by which the undamped scenario would overclose the universe, $\Omega_{\chi}^{\rm free}(X_i)/\Omega_{\rm DM}\gg 1$.

A non-quadratic vacuum potential changes the relation between $X(T_{\rm out})$ and the final abundance, but the fractional reduction of the $\chi$ energy density due to thermal damping, relative to the undamped evolution, is robust. Once suppressed, the isocurvature cannot be regenerated by the subsequent single-clock evolution.

\section{Conclusion and Outlook}
\label{s:conclusion}

Light spectator scalars are subject to stringent overclosure and CMB isocurvature bounds, which assume that the scalar evolves essentially in isolation after inflation. In this work, we have studied how thermal effects reshape the post-inflation evolution of the spectator scalar and its superhorizon perturbations. A thermal population of particles coupled to the scalar produces both a thermal effective potential and thermal damping. Both effects must be accounted for in predicting the relic abundance and perturbations of the scalar.

Our central result is that amplitude-dependent thermal damping suppresses not only the relic abundance, but also isocurvature perturbations. When thermal damping increases with the oscillation amplitude, patches with larger amplitudes are damped faster, causing sufficiently large amplitudes in different patches to converge. If, in addition, the damping at fixed amplitude grows with decreasing temperature, the amplitudes converge onto an \textit{attractor}, on which the amplitude is a unique function of the local temperature. Sustained tracking of the attractor dynamically converts an initially two-clock system into an effectively single-clock one. Thermal damping therefore relaxes the overclosure and isocurvature bounds simultaneously. In fact, we found in Eq.~\eqref{eq:punchline} that the isocurvature bound on the inflationary scale $H_{\rm inf}$ relaxes by exactly the same factor by which the overclosure bound is relaxed. This relation holds whenever the oscillation-averaged thermal damping rate scales with the oscillation amplitude $X$ as $\tilde{\Upsilon}\propto X^2$, independently of its temperature dependence.

Although we have illustrated the mechanism in a specific model, the qualitative features discussed here are not tied to it. A thermal mass and thermal damping are generic consequences of coupling a scalar to a thermal bath. Whenever thermal damping increases with the amplitude, the resulting negative feedback can lead to a dynamical attractor that sets the relic abundance and erases pre-existing isocurvature. The basin of attraction, the slope of the attractor, and its endpoint depend on the particle content and interactions, but the main conclusion, that a thermal-damping attractor relaxes both the overclosure and isocurvature bounds applies broadly.

We have not carried out a systematic study of the parameter space. Doing so for concrete, well-motivated models is a natural next step. Such a study should also address the fate of $\psi$ particles, which must disappear before BBN, the late evolution of $\chi$, and the naturalness of the bare mass of $\chi$ selected by the observed DM abundance. Concrete portals would also connect the cosmology to complementary experimental probes. Depending on the portal, the interactions responsible for thermal damping can be tested in beam-dump experiments, through stellar cooling, or in searches for ultralight dark matter. It would also be interesting to revisit scalar-induced gravitational waves and other signatures of spectator scalars in light of the enlarged parameter space.

The damping of $\chi$ heats the radiation in different patches of the universe by different amounts, since patches with larger initial amplitudes $X_i$ dissipate more energy. As in the curvaton scenario \cite{Moroi:2001ct,Lyth:2001nq,Enqvist:2001zp}, this converts part of the primordial isocurvature $I_i$ into a curvature perturbation, $\delta \zeta\sim (\rho_\chi/\rho_{\rm rad})I_i$. The ratio $\rho_\chi/\rho_{\rm rad}$ is bounded by its value at the onset of $\chi$ oscillation, $\rho_\chi/\rho_{\rm rad}\sim (X_i/M_{\rm pl})^2$. This conversion is negligible for $X_i\ll M_{\rm pl}$ in cases studied here. It can, however, become efficient and source the dominant adiabatic perturbation if $\chi$ makes up a sizable fraction of the energy density of the universe when it is damped, in which case $\chi$ acts as a curvaton. Since the attractor ties the surviving condensate to the local temperature, such a curvaton might survive as the dark matter with adiabatic perturbations. Exploring this possibility is an interesting direction for future work.

Finally, thermal damping need not always suppress isocurvature. We have found that isocurvature is damped at the rate $(1/2)\left.\partial\tilde{\Upsilon}/\partial\ln X\right|_T$. If the damping instead decreases with amplitude, patches with larger amplitudes are damped more weakly, potentially amplifying rather than erasing isocurvature. Such amplitude dependence can arise when the damping is mediated by off-shell particles or on-shell particles that are stable but pair-annihilate \cite{Tanin:2017bzm,Mukaida:2012qn,Mukaida:2013xxa}. It would be interesting to investigate this opposite case in detail. 

\section*{Acknowledgements}
We thank Itamar Allali, Dawid Brzeminski, Ngan H. Nguyen, Mudit Rai for useful discussion. E.H.T. is supported by NSF Physics Frontier Center grant PHY-2020275, NSF grant PHY-2310429, Simons Investigator Award No. 824870, the Gordon and Betty Moore Foundation Grant GBMF7946. The work of S.V. was supported by the Kavli Institute for Cosmological Physics at the University of Chicago. This work was performed in part at the Aspen Center for Physics, which is supported by National Science Foundation grant PHY-2210452.

\appendix

\section{Perturbation Equations}
\label{a:perturbationderivation}

\subsection{Derivation}
We work in Newtonian gauge with mostly minus metric signature and neglect anisotropic stress, so the two scalar metric potentials are equal,
\begin{align}
    ds^2=(1+2\Phi)dt^2-a^2(1-2\Phi)|d\mathbf{x}|^2\,,
\end{align}
with $\sqrt{-g}=a^3(1-2\Phi)$. This yields
\begin{align}
    \Box\chi=(1-2\Phi)(\ddot{\chi}+3H\dot{\chi})-4\dot{\Phi}\dot{\chi}-\frac{1}{a^2}\nabla^2\chi\,.
\end{align}
Thus the equation of motion of the scalar field $\chi=\bar{\chi}+\delta\chi$ to first order in the metric perturbation $\Phi$ is
\begin{align}
    &\ddot{\chi}-\frac{1}{a^2}\nabla^2\chi+3H\dot{\chi}+V_{\rm vac}^\prime\nonumber\\
    &=g\left<\bar{\psi}\psi\right>+4\dot{\Phi}\dot{\chi}+2\Phi(g\left<\bar{\psi}\psi\right>-V_{\rm vac}^\prime)\,. \label{eq:eomexact}
\end{align}
Close to thermal equilibrium, the response of $\psi$ can be separated into equilibrium, dissipative, and stochastic contributions. We omit the stochastic part here and return to it in Appendix~\ref{a:subleadingisocurvature}. In the local rest frame of the radiation fluid, with 4-velocity 
$u^\mu=\left(1/\sqrt{g_{00}}, v^i/a\right)$, such that $u^\mu\nabla_\mu \chi\approx (1-\Phi)\dot{\chi}$, it reads \cite{Bastero-Gil:2011rva,Mirbabayi:2022cbt}
\begin{align}
    g\left<\bar{\psi}\psi\right>=m_{\chi,\rm th}^2\left(\frac{m_\psi}{g}-\chi\right)-(1-\Phi)\Upsilon\dot{\chi}\,.
\end{align}
Its perturbation is
\begin{align}
    g\delta\left<\bar{\psi}\psi\right>=&2m_{\chi,\rm th}^2\left(\frac{m_\psi}{g}-\bar{\chi}\right)\frac{\delta T}{T}-m_{\chi,\rm th}^2\delta\chi-\bar{\Upsilon}\delta\dot{\chi}\nonumber\\
    &-\left.\frac{\partial\bar{\Upsilon}}{\partial\bar{\chi}}\right|_T\dot{\bar{\chi}}\delta\chi-\left.\frac{\partial\bar{\Upsilon}}{\partial\ln T}\right|_{\bar{\chi}}\dot{\bar{\chi}}\frac{\delta T}{T}+\bar{\Upsilon}\dot{\bar{\chi}}\Phi\,.
\end{align}
The evolution equation for the superhorizon perturbation $\delta\chi$, Eq.~\eqref{eq:perturbationevolution} is obtained by perturbing Eq.~\eqref{eq:eomexact}, substituting the above, keeping the first-order terms in $\Phi$, $\delta\chi$, and $\delta T$, and taking the superhorizon $k\rightarrow 0$ limit. Here we take $V_{\rm vac}^\prime=V_{\rm vac}^{\prime\prime}\bar{\chi}$ and drop terms suppressed by $m_\psi/(gX)$.

\subsection{Phase and Amplitude}
Any superhorizon perturbation $\delta\chi$ can be decomposed into a $\propto\cos\theta$ piece (in phase with $\bar{\chi}=X\cos\theta$) and a $\propto -\sin\theta$ piece  (\textit{nearly} in  phase with $\dot{\bar{\chi}}=-X\dot{\theta}\sin\theta+\dot{X}\cos\theta$),
\begin{align}
    \delta \chi=X\left(\Delta_X\cos\theta-\Delta_\theta\sin\theta\right)\,,
\end{align}
where $\Delta_X=\delta X/X$, $\Delta_\theta=\delta\theta$, and we assume that the envelopes of both modes vary slowly, $|\dot{\Delta}_X|/\Delta_X,|\dot{\Delta}_\theta|/\Delta_\theta\ll M_\chi$. Inserting the above ansatz into the $\delta\chi$-evolution equation, multiplying by either $\cos\theta$ or $\sin\theta$, and then time averaging 
using $\left<\cos^2\theta\right>=\left<\sin^2\theta\right>=1/2$, $\left<\sin\theta\cos\theta\right>=0$, $\left<\sin^2\theta\cos^2\theta\right>=1/8$, $\left<\cos^4\theta\right>=3/8$, $\left<\sin\theta\cos^3\theta\right>=\left<\sin^3\theta\cos\theta\right>=0$
yields
\begin{align}
    \dot{\Delta}_X+\frac{1}{2}\left.\frac{\partial\tilde{\Upsilon}}{\partial \ln X}\right|_T\Delta_X&=- 
    \frac{\left<S\sin\theta\right>}{M_\chi X}\nonumber\\
    &=2\dot{\Phi}-\frac{1}{2}\tilde{\Upsilon}\Phi-\frac{1}{2}\left.\frac{\partial\tilde{\Upsilon}}{\partial\ln T}\right|_{X}\frac{\delta T}{T}\,,\\
    \dot{\Delta}_\theta&=-\frac{\left<S\cos\theta\right>}{M_\chi X}\nonumber\\
    &=M_\chi\Phi+\frac{m_{\chi,\rm th}^2}{M_\chi}\frac{\delta T}{T}\,,
\end{align}
Note that the phase perturbation $\Delta_\theta$ is not thermally damped.

\subsection{Adiabatic and Isocurvature}
The $\cos\theta$-$\sin\theta$ decomposition discussed above is convenient for deriving the time-averaged evolution of $\delta\chi$ modes, but it does not coincide with the cosmological adiabatic-isocurvature split.  

An intuitive way to separate adiabatic and isocurvature modes is to picture it geometrically \cite{Gordon:2000hv}. We depict the background trajectory $\bar{\chi}(t)$ in the complex plane by writing $\bar{\chi}=\text{Re}\left[Xe^{i\theta(t)}\right]$. Adiabatic and isocurvature perturbations are, respectively, those along and orthogonal to the $\bar{\chi}(t)$ trajectory on the complex plane. If $X$ were constant, the trajectory would be circular, and it is then clear that the phase perturbation ($\delta\theta=\Delta_\theta$) is adiabatic and the amplitude perturbation ($\delta X/X=\Delta_X$) is isocurvature. For nonzero $\dot{X}$, the trajectory becomes a spiral and its tangent has both angular and radial components. The phase perturbation ($\delta\theta=\Delta_\theta$) remains, to leading order, the adiabatic, tangent mode with a subleading, $\mathcal{O}[(|\dot{X}|/X)/M_\chi)]$ isocurvature contamination. The amplitude perturbation, on the other hand, receives an adiabatic contribution that is not small. Since $|\dot{X}/X|\ll M_\chi$ implies that the background motion is dominantly in the phase direction, even a small tilt in the $\bar{\chi}$ trajectory away from the phase direction contributes an adiabatic piece $\dot{X}\delta t/X$ to the radial mode $\Delta_X=\delta X/X$. In a radiation-dominated universe, the time shift is $\delta t=\delta T/\dot{T}$. The isocurvature is the residual amplitude that is not explained by a time shift, $\Delta_X-\dot{X} (\delta T/\dot{T})/X$. 

Therefore one can, up to an arbitrary normalization, choose the following adiabatic ($A$) and isocurvature ($I$) modes 
\begin{align}
    A\equiv \Delta_\theta,\quad I\equiv \Delta_X-\xi\frac{\delta T}{T}\,,
\end{align}
which evolve as
\begin{align}
    \dot{A}=&M_\chi\Phi+\frac{m_{\chi,\rm th}^2}{M_\chi}\frac{\delta T}{T}\,,\\
    \dot{I}+\frac{1}{2}\left.\frac{\partial\tilde{\Upsilon}}{\partial \ln X}\right|_TI=&2\dot{\Phi}-\frac{d}{dt}\left(\xi\frac{\delta T}{T}\right)-\frac{1}{2}\tilde{\Upsilon}\Phi\nonumber\\
    &-\left(\frac{1}{2}\left.\frac{\partial\tilde{\Upsilon}}{\partial\ln X}\right|_T\xi+\frac{1}{2}\left.\frac{\partial\tilde{\Upsilon}}{\partial\ln T}\right|_X\right)\frac{\delta T}{T}\nonumber\\
    \approx&-\left[\dot{\xi}+\tilde{\Upsilon}\left(\xi-\xi_\Upsilon\right)\right]\frac{\delta T}{T}\,.
\end{align} 
The amplitude dependence of $\tilde{\Upsilon}$ is what damps the $I$. For $\tilde{\Upsilon}\propto X^2$, $\partial\tilde{\Upsilon}/\partial\ln X=2\tilde{\Upsilon}$. In the last step of the $\dot{I}$ equation, we assume $\tilde{\Upsilon}\propto X^2$ and the spectator limit (valid deep in radiation domination) where the radiation continuity equation $\delta\dot\rho_r+4H\delta\rho_r=4\dot{\Phi}\rho_r$ reduces to $d(\delta T/T)/dt=\dot{\Phi}$, with $\dot\Phi\approx 0$ and $\delta T/T\approx -\Phi/2$; we also used Eq.~\eqref{eq:xiups}.

The structure of these equations reflects Weinberg's adiabatic theorem \cite{Weinberg:2003sw}. The phase equation is solved exactly by the time shift $\delta t=\Phi t=-(2\delta T/T)t$ for $M_{\chi}\approx m_{\chi,\rm th}$. Consequently the phase isocurvature $\Delta_\theta-M_\chi \delta t$ is conserved on superhorizon scales. The source of $I$ vanishes on any background solution (see Section~\ref{s:perturbation}), so an initially zero $I$ will remain zero. The gravitational source $2\dot{\Phi}$ of $I$ switches on only as a mode crosses the horizon, where $\dot{\Phi}\sim (k/aH)^2H\Phi$. This generates the blue-tilted isocurvature of order $(k/aH)^2$ correlated with the adiabatic one; see Appendix~\ref{a:subleadingisocurvature}.

\section{$\chi$-Radiation Relative Isocurvature}
\label{a:relativeisocurvature}
The coarse-grained continuity equations for the oscillation-averaged energy densities $\rho_\chi=M_\chi^2X^2/2$ and $\rho_r\propto T^4$ of the $\chi$ and radiation fluids read
\begin{align}
    \dot{\rho}_\chi+3H(1+w_\chi)\rho_\chi&=Q_{\chi}\,,\\
    \dot{\rho}_{r}+4H\rho_{r}&=Q_{r}\,,
\end{align}
where
\begin{align}
    w_\chi=-\frac{1}{3H}\frac{\dot{M}_\chi}{M_\chi},\quad Q_r=-Q_\chi=\tilde{\Upsilon}\rho_\chi\,.
\end{align}
We work in the spectator limit, in which $\rho_\chi\ll \rho_r$ and $\tilde{\Upsilon}\rho_\chi\ll 4H\rho_r$, and so $\chi$ does not backreact appreciably on the radiation and the expansion.  The oscillation-averaged density perturbations are
\begin{align}
    \frac{\delta\rho_\chi}{\rho_\chi}&=2\frac{\delta X}{X}+\frac{\delta (M_\chi^2)}{M_\chi^2}=2I+2\left(\xi+\frac{m_{\chi,\rm th}^2}{M_\chi^2}\right)\frac{\delta T}{T}\,,\\
    \frac{\delta\rho_{\rm rad}}{\rho_{\rm rad}}&=4\frac{\delta T}{T}\,.
\end{align}
We define the curvature perturbation on uniform-$\rho_\alpha$ slices $\zeta_\alpha$ as the $e$-fold shift between flat slicing and uniform-$\rho_\alpha$ slicing $\zeta_\alpha=-\Phi-H \delta\rho_{\alpha}/\dot{\rho}_\alpha$. The total curvature perturbation is the weighted sum $\zeta=\sum_\alpha (\dot{\rho}_\alpha/\dot{\rho})\zeta_\alpha$, which in the limit that the energy density of $\chi$ is negligible, reduces to $\zeta\approx \zeta_r$. The relative isocurvature between $\chi$ and radiation is $S_{\chi r}\equiv3(\zeta_\chi-\zeta_r)$ and can be expressed as
\begin{align}
    S_{\chi r}&=-3H\left(\frac{\delta\rho_\chi}{  \dot{\rho}_\chi}-\frac{\delta\rho_r}{  \dot{\rho}_r}\right)=\frac{3I}{\xi+m_{\chi,\rm th}^2/M_\chi^2}\,,
\end{align}
where we have used $\dot{\rho}_\chi/\rho_\chi=-2H[m_{\chi,\rm th}^2/M_\chi^2+\xi]$. This further reduces to $S_{\chi r}=3I/(1+\xi)$ when the potential of $\chi$ is thermal-mass dominated.

In the interacting-fluid formalism of Ref.~\cite{Malik:2004tf}, the erasure of $S_{\chi r}$ is driven by the perturbed energy transfer, $\delta Q_\chi=-\delta(\tilde{\Upsilon}\rho_\chi)$, which has a non-adiabatic component stemming from the $X$ dependence of $\tilde{\Upsilon}$. If $\tilde{\Upsilon}$ depended on $T$ alone, $\delta Q_\chi$ would be completely adiabatic.

\section{Subleading Isocurvature Sources}
\label{a:subleadingisocurvature}

We find in the main text that thermal-damping attractor tracking exponentially suppresses the spectator isocurvature mode, resulting in a nearly adiabatic final perturbation. This conclusion holds in the separate-universe (deep superhorizon) limit and after neglecting stochastic thermal fluctuations of the radiation. Spatial gradients and thermal noise can produce additional isocurvature perturbations, but both have spectra that are strongly blue and are negligible on CMB scales. Below we denote the temperature below which thermal damping becomes negligible as $T_{\rm out}$.

\subsection{Gradient Effects}

Reinstating the gradient terms $(k/a)^2\delta\chi$ and $\dot{\Phi}\sim (k^2/a^2H^2)H\Phi$ in Eq.~\eqref{eq:perturbationevolution} produces corrections to resulting $\delta\chi$. These effects lead parametrically to \cite{Holst:2023msh,Stebbins:2023wak}
\begin{align}
    S_{\chi r}\sim \left(\frac{k}{a H}\right)^2 \zeta\,.
\end{align}
The resulting isocurvature is correlated with the curvature perturbation $\zeta$. The mode crossing the horizon at the end of tracking is
$k_{\rm out}=\left.aH\right|_{T_{\rm out}}$ which evaluates to $k_{\rm out}=1\times 10^{4}\text{ Mpc}^{-1}\left(T_{\rm out}/\MeV\right)$,
where we took $g_*(T_{\rm out})=10.75$. Note that $T_{\rm out}\gtrsim\MeV$ by construction. In the extreme case where tracking persists until the onset of BBN, $T_{\rm out}\sim 1\MeV$, the modes
$k\sim 2\times 10^4\text{ Mpc}^{-1}$ remain superhorizon throughout the damping epoch. At the CMB pivot scale $k_*=0.05\text{ Mpc}^{-1}$, for instance, one has $(k_*/k_{\rm out})^2\sim 2\times 10^{-11}(T_{\rm out}/\MeV)^{-2}$. Therefore the corresponding suppression  in the isocurvature power spectrum, $(k_*/k_{\rm out})^4$, is typically more severe than $10^{-22}$, far beyond the current observational reach. Nevertheless it is a prediction of our scenario that could serve as an interesting target for probes of higher-$k$ modes, such as Ref.~\cite{Graham:2024hah}.

\subsection{Thermal Noise}

The fluctuation-dissipation relation implies that thermal damping must be accompanied with stochastic fluctuations. The perturbation equation Eq.~\eqref{eq:perturbationevolution} should therefore include a thermal noise source $\eta$. A cosmological scale typically contains an enormous number of independent thermal-correlation domains, which add up incoherently and largely cancel, up to Poisson mis-cancellations. More formally, at scales larger than the thermal correlation length, the noise is approximately Gaussian with correlator $\left<\eta(t,\mathbf{x})\eta(t',\mathbf{x}')\right>=(2\Upsilon T/a^3)\delta(t-t')\delta^{(3)}(\mathbf{x}-\mathbf{x}')$, which corresponds to white noise $\left<\eta_k(t)\eta_{k'}(t')\right>=(2\Upsilon T/a^3)(2\pi)^3\delta^{(3)}(\mathbf{k}+\mathbf{k}')\delta(t-t')$ in the $k$-space. Its effect is highly suppressed by the smallness of $k/aH$ on observationally relevant scales \cite{Bae:2025vkn}.

\section{Models}
\label{a:models}

We have parametrized the thermalization rate of $\psi$ as $\Gamma_{\rm th}=\kappa T^n$. If the scalar primarily perturbs the number density of $\psi$ then the $\psi$ thermalization rate $\Gamma_{\rm th}$ is set by number-changing processes that restore chemical equilibrium of $\psi$ with the primordial plasma. If it instead mainly distorts the momentum distribution at fixed number density, $\Gamma_{\rm th}$ is set by processes that restore kinetic equilibrium. In either case, the relevant rate is the slowest one needed to restore equilibrium.

The $n=1$ case can be realized, for instance, with a light vector mediator ($m_{V}\ll T$) coupled to both $\psi$ ($V_\mu \bar{\psi}\gamma^\mu\psi$) and SM fermions ($V_\mu \bar{\psi}_{\rm SM}\gamma^\mu\psi_{\rm SM}$); the same model realizes $n=5$ if the mediator is heavy ($m_{V}\gg T$). If the dominant number-changing process is $2\leftrightarrow 2$ annihilation $\psi\bar{\psi}\leftrightarrow\text{SM}\,\text{SM}$, then the thermalization rate is parametrically $\Gamma_{\rm th}\sim n_\psi\sigma_{\psi\leftrightarrow\text{SM}}\sim T^3\sigma_{\psi\leftrightarrow\text{SM}}$. If thermalization is mediated by an operator of mass dimension $d$, suppressed by $\Lambda^{4-d}$, dimensional analysis gives $\Gamma_{\rm th}\sim T^{2d-7}/\Lambda^{2d-8}$, which corresponds to $n=2d-7$ and $\kappa\sim \Lambda^{8-2d}$. For $d=4,5,6,7,\ldots$, this gives $n=1, 3,5,7,\ldots$; $d=4$ is the light vector mediator case above, $d=5$ includes the fermionic Higgs portal $(\bar{\psi}\psi)(H^\dagger H)/\Lambda$, and $d=6$ includes four-fermion operators and the heavy limit of the vector mediator above. Thermalization via relativistic $2\rightarrow2$ annihilation therefore naturally produces positive odd values of $n$. The case $n=-1$ can be realized with $S\bar{\psi}\psi$, for which the (inverse-)decay rate is $\Gamma_{\rm th}\propto M_\psi^2/T$. 

Some of these example thermalization rates $\Gamma_{\rm th}$ introduce dependence on $M_\psi$, and hence on $X$. In those cases the attractor analysis must be redone with the modified $X$ dependence.

\clearpage
\bibliography{references}

\end{document}